\documentclass[aps,showkeys,showpacs,twocolumn]{revtex4-1}

\usepackage{psfrag}
\usepackage{graphicx}
\usepackage{amsmath}
\usepackage{hyperref}
\usepackage{color}
 
\begin{document}

\title{Bounded confidence model: addressed information maintain diversity of opinions}

\author{K.~Malarz}
\homepage{http://home.agh.edu.pl/malarz/}
\email{malarz@agh.edu.pl}
\affiliation{
AGH University of Science and Technology, Faculty of Physics and Applied Computer Science, al. Mickiewicza 30, 30-059 Krakow, Poland
}

\author{K.~Ku{\l}akowski}
\email{kulakowski@novell.ftj.agh.edu.pl}
\affiliation{
AGH University of Science and Technology, Faculty of Physics and Applied Computer Science, al. Mickiewicza 30, 30-059 Krakow, Poland
}
 
\date{\today}
 
\begin{abstract}
A community of agents is subject to a stream of messages, which are represented as points on a plane of issues. Messages are sent by media and by agents themselves. Messages from media shape the public opinion. They are unbiased, i.e. positive and negative opinions on a given issue appear with equal frequencies. In our previous work, the only criterion to receive a message by an agent is if the distance between this message and the ones received earlier does not exceed the given value of the tolerance parameter.  Here we introduce a possibility to address a message to a given neighbour. 
We show that this option reduces the unanimity effect, what improves the collective performance.
 \end{abstract}
 
 

\maketitle
  
\section{\label{sec-itro} Introduction}

Problems of communication are at the centre of attention of several branches of sociology \cite{oxf}, with some overlaps from psychology to telecommunication. There, statistical physics has also a bridgehead \cite{sta,rev}, where ideas like agent simulations \cite{ags}, social networks \cite{new} and evolutionary games \cite{wei} are as common as non-equilibrium processes \cite{dick} and Monte Carlo simulations \cite{mcr}.
Despite this interest, a split remains between social sciences and sociophysics \cite{me,mem}, because different aspects of reality are qualified as interesting by researchers on both sides. It is worthwhile, then, for a sociophysicist to investigate a problem formulated by a social scientist; in this way, the above split can be at least reduced. Here we are interested in the problem how messages are received and accepted, as formulated by John Zaller \cite{zal}---the equations of Zaller are briefly summarised also in 2-nd section of \cite{ja}. Although the mathematical formalism we use here differs from the original one, the core of the social process remains---we hope---unchanged.

\begin{figure}
\psfrag{mu}{$\mu$}
\psfrag{A1}{$\mathbf{A_1}$}
\psfrag{A2}{$\mathbf{A_2}$}
\psfrag{A3}{$\mathbf{A_3}$}
\psfrag{A4}{$\mathbf{A_4}$}
\psfrag{A10}{$A_1^0$}
\psfrag{A20}{$A_2^0$}
\psfrag{A30}{$A_3^0$}
\psfrag{A40}{$A_4^0$}
\psfrag{A11}{$A_1^1=M^1$}
\psfrag{A12}{$A_1^2=M^2$}
\psfrag{A13}{$A_1^3=M^8$}
\psfrag{A21}{$A_2^1=M^5$}
\psfrag{A22}{$A_2^2=M^7$}
\psfrag{A31}{$A_3^1=M^4$}
\psfrag{A32}{$A_3^2=M^9$}
\psfrag{M3}{$M^3$}
\psfrag{M6}{$M^6$}
\psfrag{M10}{$M^{10}$}
\psfrag{M11}{$M^{11}$}
\psfrag{M12}{$M^{12}$}
\psfrag{mu}{$\mu$}
\includegraphics[width=0.45\textwidth]{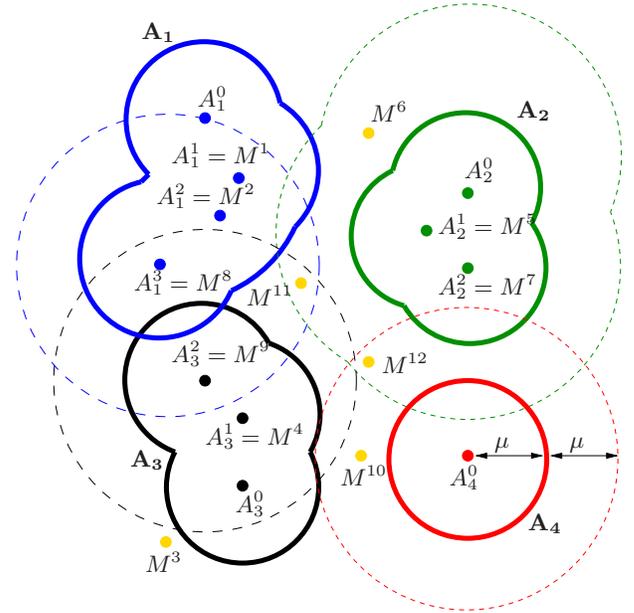}
\caption{\label{fig-1} (Colour online). Scheme of message plane. See Sec. \ref{sec-care} for detailed description.}
\end{figure}

In this process, a community is subject to a stream of messages from media. They are noticed or not, depending on how their political content fits 
individual profiles of the receivers; further, they are accepted or not on a similar basis. The model description by Zaller \cite{zal} was reformulated
in \cite{ja} in the spirit of the bounded confidence model \cite{def}, where messages are represented as points in the plane of issues. In the next step, 
communication between agents was introduced to the model \cite{jas}. This generalisation mate the approach even more close to the bounded confidence model, then it is justified to term it as the Zaller--Deffuant model. In both works \cite{ja,jas}, a community was subject to a stream of messages, where positive and negative opinions on a given issue appear with equal frequencies. The proper outcome of opinions should then be ``I don't know'', and the
quality of the social performance can be measured by an evaluation of the variance $\sigma ^2$ of the opinion distribution. A large variance could be interpreted as many extreme opinions; in this case, the performance is poor. On the contrary, a small variance means that most people remain undecided. In \cite{ja}, where the communication between agents was absent, the variance was found to be anti-correlated with the tolerance parameter $\mu$, which measured an individual ability to receive/accept distant messages. In \cite{jas}, an intensive exchange of messages between agents was demonstrated to lead to unanimity. This option was found to deteriorate the collective performance, by producing agents who are strongly convinced in their opinions. Simply, the probabilities that an agent answers `Yes' and `No' were always either one and zero or zero and one, but never a half and a half. This was a consequence of an effective attraction between agents on the plane of issues \cite{def}. Paradoxically, for large values of the tolerance parameter $\mu$ the attraction was even more effective, and the deterioration of the performance was even stronger, than for small $\mu$.

Below we consider a modification of the process. Previously \cite{jas}, the only criterion to receive a message by an agent is if the distance between this message and the ones received earlier does not exceed the given value of the tolerance parameter. Now, agents address their messages to those neighbours which are most close in the plane of issues. There, the distance is between most close opinions. Moreover, the tolerance parameter for the interpersonal messages is assumed to be twice larger than its value for the messages from media. The goal of this paper is to demonstrate, that this individualised way of communication destroys the unanimity. As a consequence, the variance of opinions $\sigma ^2$ decreases monotonously with the mean value of the tolerance parameter $\mu$.

\section{\label{sec-care} Calculations and results}

The message space is a square $\mathcal{S}=\{(x,y):-1\le x\le+1,-1\le y\le+1\}$.
Initially, $\mathcal{N}$ agents are randomly distributed inside this square at positions $A_i^0=(x_i^0,y_i^0)\in\mathcal{S}$ (for $i=1,\cdots,\mathcal{N}$).
This set of agents is exposed to the stream of subsequent $\mathcal{M}$ external messages, which appear on the message space at randomly selected positions $M^t=(x^t,y^t)\in\mathcal{S}$ (for $t=1,\cdots,\mathcal{M}$).
Agent $i$ accepts information $A_i^\tau=M^t$ if any of his/her so far accepted messages $\mathbf{A}_i=\{A_i^0, A_i^1, \cdots, A_i^{\tau(i)-1} \}$ is closer than $\mu$ to $M^t$. 

Between each two external messages, approximately ten messages are sent by randomly selected agents to their nearest neighbours, in the same way as in \cite{jas}. To complete this, a nearest neighbour $n(i)$ for each agent $i$ is detected. As noted above, distances are compared between nearest messages previously accepted by given agents. Then each agent $i$ tries to send to his/her nearest neighbour $n(i)$ this of so far accepted messages $A_i^j$ $(0\le j< \tau(i))$ which is closest to the neighbour's information $\mathbf{A}_{n(i)}=\{ A_{n(i)}^0, A_{n(i)}^1, \cdots, A_{n(i)}^{\tau(n(i))-1} \}$.
If this selected and sent message $A_i^j$ is closer than $2\mu$ to the set $\mathbf{A}_{n(i)}$ then it is accepted by $n(i)$ as his/her $(\tau+1)$-th message $A_{n(i)}^\tau=A_i^j$, where $\tau=\tau(n(i))$.

In Fig. \ref{fig-1} the schematic sketch of the messages space with initial agents distribution, their set of accepted messages and incoming messages are presented.
The initial agents' positions are $A_1^0$, $A_2^0$, $A_3^0$ and $A_4^0$.
A dozen of subsequent messages appear at the positions $M^1, M^2, \cdots, M^{11}, M^{12}$.
Among them messages $M^{3,6,10-12}$ were neglected by all agents.
The subsequent sets of messages $(M^1,M^2,M^8)$, $(M^5,M^7)$ and $(M^4,M^9)$ were accepted by agents $i=1,2$ and 3, respectively.
The solid lines represent the borders of agent's acceptance area for incoming messages. The dashed lines show the borders for interpersonal interaction (information exchange) among the nearest neighbours.
Messages $A_1^3$ and $A_3^2$ will be shared among agents $i=1,3$ as soon as the message $M^9$ arrives.

Similarly to the procedure described in Ref. \cite{tai} we evaluate the normalised probability $p_i$ of positive answers to some questions asked to $i$-th agent as
\begin{equation}
p_i=\frac{\sum_{j=1}^{\tau(i)} x^j_i H(x^j_i)}{\sum_{j=1}^{\tau(i)} |x^j_i|},
\end{equation}
where $x^j_i$ is the $x$-th coordinate of the $j$-th message received by $i$-th agent, and $H(x)$ is Heaviside step function
\[
H(x)=
\begin{cases}
 1 & \iff x\ge 0,\\
 0 & \iff x<0.
\end{cases}
\]

\begin{figure}
\psfrag{mu}{$\mu=$}
\psfrag{P(p)}{$P(p_i)$}
\psfrag{p}{$p_i$}
\includegraphics[width=0.45\textwidth]{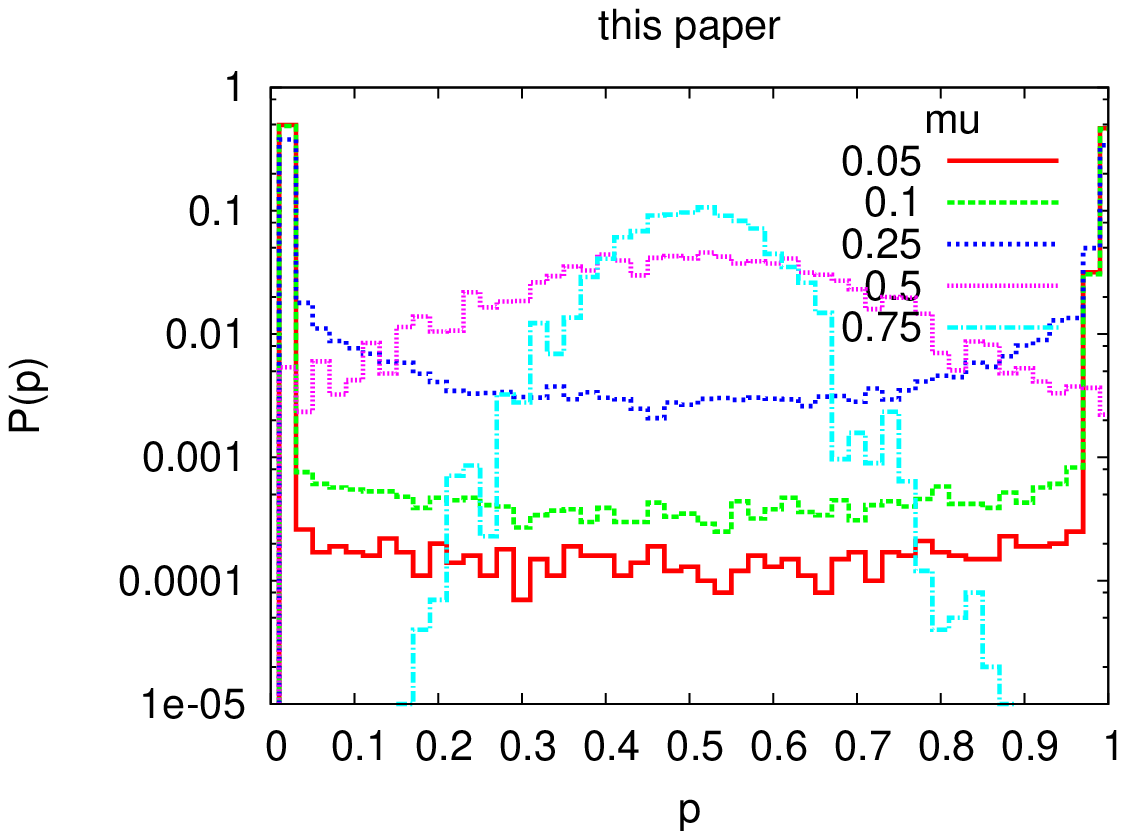}\\
\includegraphics[width=0.45\textwidth]{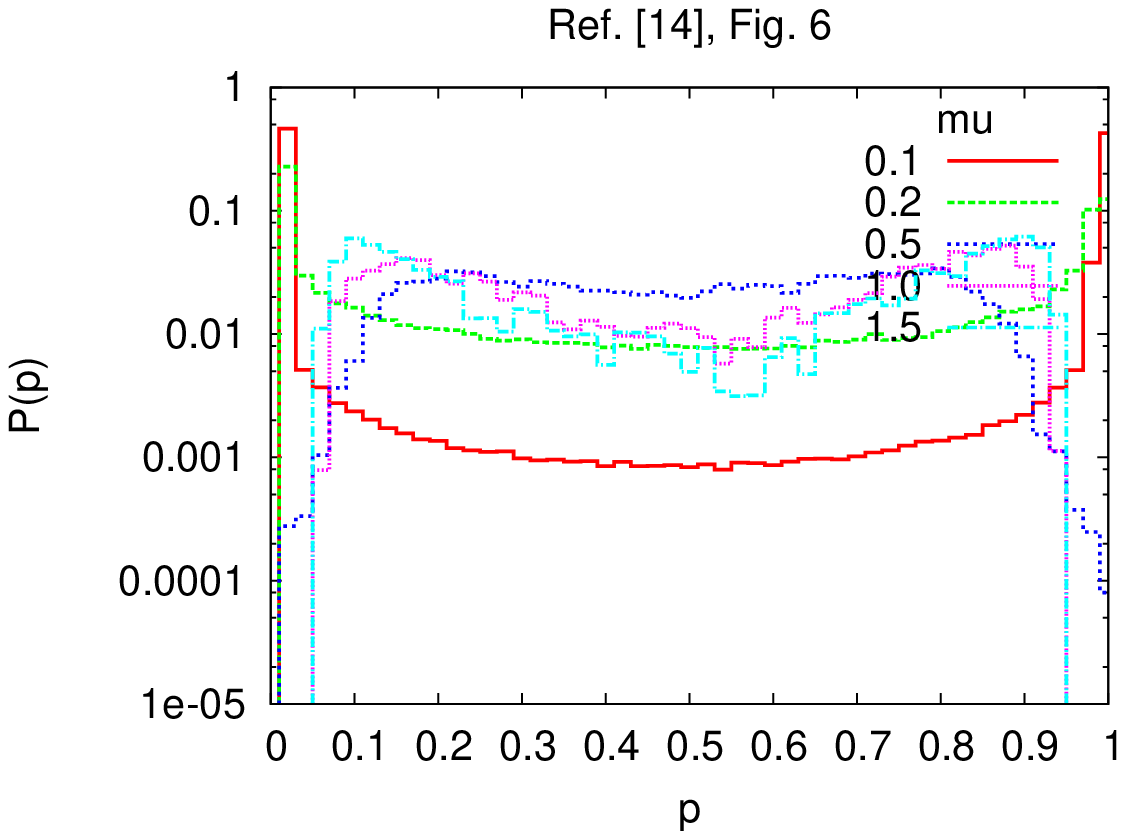}
\caption{\label{fig-2} Histogram $P(p_i)$ of a relative probabilities $p_i$ for various values of tolerance parameter $\mu$. The initial number of agents $\mathcal{N}=10^3$ is a subject to a stream of $\mathcal{M}=100$ messages. The results are averaged over $N_{\text{run}}=100$ independent simulations.}
\end{figure}

\begin{figure}
\psfrag{mu red, 1.5mu green}{$\mu$ (red), $1.5\mu$ (green)}
\psfrag{Var(pi)}{$\sigma^2(p_i)$}
\includegraphics[width=0.45\textwidth]{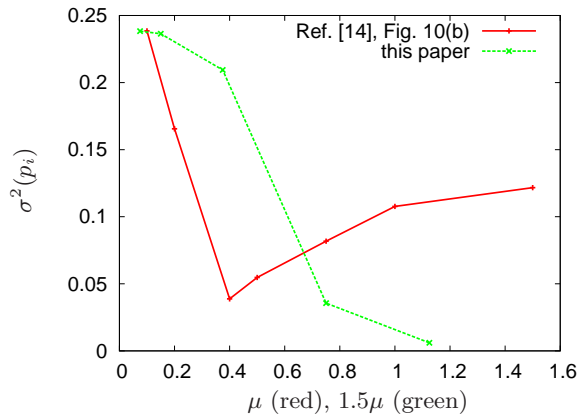}
\caption{\label{fig-3} Variance $\sigma^2(p_i)$ for various values of a tolerance parameter $\mu$.  The values of $\mu$ for current version of simulation are enlarged by a factor 3/2 as an average of external ($\mu$) and internal $(2\mu)$ tolerance parameter. $\mathcal{N}=10^3$, $\mathcal{M}=100$, $N_{\text{run}}=100$.}
\end{figure}

In Fig. \ref{fig-2} the histogram $P(p_i)$ of probabilities $p_i$ for $\mathcal{N}=10^3$ agents exposed to a stream of subsequent $\mathcal{M}=100$ messages and for various values of the tolerance parameter $\mu$ is presented.
The results are averaged over $N_{\text{run}}=100$ independent simulations. Note that while for one simulation the state of unanimity is characterized by a small variance, the variance averaged over populations with unanimous and extreme opinions is large.
For comparison the same histogram for agents whose tolerance parameter $\mu$ is the same for external and interpersonal communication is replotted.
In Fig. \ref{fig-3} the variance $\sigma^2(p_i)$ for distribution  $P(p_i)$ given in Fig. \ref{fig-2} is presented. The values of $\mu$ for current version of simulation are enlarged by a factor 3/2 as an average of external ($\mu$) and internal $(2\mu)$ tolerance parameter.

\section{\label{sec-disc} Discussion}

While many people believe that the content of laws of physics is free of values, the situation in sociology is much less clear \cite{wilmay}. {\em Toutes proportions gard\'ees}, we can repeat the question of values, regarding our results. The unanimity can be treated as a valuable state, where long discussions do not destroy a coherent action. However, the condition of unanimity can also be a threat for individual freedom. As we see, values are to be selected on basis of individual experience and of demands of the situation.

In a recent work on a similar model, we evaluated the statistical meaningfulness of communities as dependent on the tolerance parameter $\mu$ \cite{tai}. We found, that the community structure is most meaningful when the overall communication is weak; in this case social connections are rare but once some exist, they strongly determine the local behaviour. These results are parallel and supplementary to the results obtained here. As an integrated picture, an opposition emerges: a common and uniform communication against a local one, unanimity against diversity of opinions, loose contacts with everybody or well-defined social clusters. The role of the parameter of tolerance remains ambiguous. When an interpersonal communication is absent, the tolerance improves understanding of messages from media \cite{ja}, but in the presence of communication it can lead to unanimity around a random opinion \cite{jas}. Our main conclusion here is that individually addressed messages maintain the diversity. 

\begin{acknowledgments}
One of authors (K.K.) acknowledges inspiring discussions with Sylvie Huet.
Partially supported by the Polish Ministry of Science and Higher Education and its grants for scientific research and within the FP7 project SOCIONICAL (grant No. 231288).
The numerical calculations were carried out in the Academic Computer Centre CY\-F\-RO\-NET\---AGH (grant No. MEiN/SGI3700/\-AGH/\-024/2006).
\end{acknowledgments}

\end{document}